\documentclass[aip,cha,preprint]{revtex4-1}
\usepackage{lineno}
\usepackage{graphicx}
\usepackage{natbib}
\usepackage{color}

\begin{document}
%\linenumbers

\title{Shape analysis using fractal dimension: a curvature based approach}

\author{Andr\'e R. Backes}
	\email{backes@facom.ufu.br}
\affiliation{Faculdade de Computa\c{c}\~{a}o, Universidade Federal de Uberl\^{a}ndia\\ Av. Jo\~{a}o Naves de \'{A}vila, 2121, 38408-100 Uberl\^{a}ndia, Minas Gerais, Brazil}

\author{Jo\~{a}o B. Florindo}
	\email{jbflorindo@ursa.ifsc.usp.br}
\affiliation{Instituto de F\'{i}sica de S\~{a}o Carlos (IFSC)\\ Universidade de S\~{a}o Paulo\\ Av.  Trabalhador S\~{a}o Carlense, 400,
CEP 13560-970\\S\~{a}o Carlos, S\~{a}o Paulo, Brasil} 

\author{Odemir M. Bruno}
              \email{bruno@ifsc.usp.br}
\affiliation{Instituto de F\'{i}sica de S\~{a}o Carlos (IFSC)\\ Universidade de S\~{a}o Paulo\\Av.  Trabalhador S\~{a}o Carlense, 400,
CEP 13560-970\\S\~{a}o Carlos, S\~{a}o Paulo, Brasil}       

%\date{}

\begin{abstract}
The present work shows a novel fractal dimension method for shape analysis. The proposed technique extracts descriptors from the shape by applying a multiscale approach to the calculus of the fractal dimension of that shape. The fractal dimension is obtained by the application of the curvature scale-space technique to the original shape. Through the application of a multiscale transform to the dimension calculus, it is obtained a set of numbers (descriptors) capable of describing with a high precision the shape in analysis. The obtained descriptors are validated in a classification process. The results demonstrate that the novel technique provides descriptors highly reliable, confirming the precision of the proposed method.
\end{abstract}

\keywords{
Shape Analysis, Curvature, Complexity, Multi-scale Fractal Dimension.
}

\maketitle

\section{Introduction}

One of the most important research areas in pattern recognition and image analysis is the analysis of shapes. Shapes are capable of describing an object by preserving the most relevant information about it, simplifying its aspect. This enables us to use computational algorithms to solve important problems in computational vision \cite{Loncaric98}.

Particularly, the literature presents various methods aiming to provide a single number or a set of them, which can be used to describe the essential characteristics of the shape. These numbers are called shape descriptors. Among the methods for the obtainment of these descriptors, one that has a growing number of works showed in the literature is the fractal descriptors \cite{Plotze-2005}.

Formally defined in \cite{M68}, the fractal geometry studies mathematical entities, which cannot be perfectly modelled and described by the conventional Euclidean geometry. In the Nature, we can find a number of these objects, such as the sky clouds, the trees, the river streams, etc. Mandelbrot \cite{M68} presents fractal objects as complex structures characterized mainly by the self-similarity, i.e., each part of the object is a copy in reduced scale of the original object. In authentic fractal objects, the self-similarity property is seen in infinite levels of scale. 

In his study about fractal geometry, Mandelbrot \cite{M68} showed more formally the concept of fractal dimension and the possibility of its use as a measure to describe fractal objects. In \cite{Carlin00}, it is still demonstrated that the fractal geometry and, consequently, the fractal dimension concept, could be extended to objects that are not necessarily true fractal entities, but, which demonstrate characteristics of fractals as the self-similarity (although to a limited degree). With the advent of computers and the representation of natural objects by digital images, the study of fractal methods provided an important source of descriptors for objects in natural scenes. Recently, we can find several works in the literature using the fractal dimension as a descriptor of objects in applications at different areas of the science \cite{BPVA09,JGCSZ09}.

In the specific case of shape analysis, \cite{Plotze-2005} presents a novel approach of fractal descriptors, where the concept of Multi-scale Fractal Dimension (MFD) is defined. In that approach, the fractal dimension is not used as a unique descriptor of an object but the dimension is calculated at different levels of observation (from a micro to a macroscopic scale). Thus, a set of numbers (each one corresponding to a different observation scale) is generated to compose a vector of fractal descriptors. This approach has demonstrated to represent the shape with more richness than the use of a simple fractal dimension descriptor and many other works have followed the same approach and showed interesting results in pattern recognition problems \cite{Plotze-2005,Torres2003,BackesMVA}.

The present work proposes a novel MFD approach based on the fractal dimension calculus method developed in \cite{BFB09}. In \cite{BFB09} the fractal dimension is calculated by the multi-scale approach called Curvature Space Scale (CSS) \cite{costa-2000}, in which the fractal dimension is calculated by the determination of the value of geometric curvature of the shape at a varying scale of observation. The proposal of this work is to use the set of curvature measures as a set of MFD descriptors for the shape. The method is validated in a task of classification of object shapes over a dataset whose classes are previously known.

The next sections show the main mathematical concepts used in this work, i.e., curvature and fractal dimension. It also describes the CSS technique and how it can be used for the calculus of fractal dimension. The following section describes the Multi-scale Fractal Dimension method. An experiment to validate our approach is set on the posterior section. The next section presents the results and discussion and the last one does the conclusions of the work.

\section{Shape and Fractal Dimension}

The use of shapes to characterize objects in a scene constitutes an important branch of pattern recognition \cite{Loncaric98}. The shape of an object can be described as a binary image representing the silhouette of the analyzed object. It contains the main information relative to the original object and shows less complexity than the representation in the original image.

In pattern recognition applications, it is interesting to represent a shape by a set of numerical descriptors, i.e., a feature vector. The literature shows various techniques to generate these descriptors. Particularly, one interesting shape descriptor is the fractal dimension.

The concept of fractal objects dates back to the $19^{th}$ century. However, Mandelbrot \cite{M68} is the first known author to show a more formally detailed study. Fractals are objects with infinite complexity, generated from simple Euclidean geometric entities and following, in most cases, a simple rule of construction. They are also characterized by their self-similarity, i.e., if one takes a small piece of the object, it can notice that this piece is an approximated copy, reduced in scale, of the original object. In true fractal objects, this property can be observed at all levels of scale (infinitely).

Mandelbrot noticed in his work that the fractals were an interesting tool to model objects found in the nature. In fact, we can easily observe that there are various objects around us, which are composed by the simple repetition of a pattern. Thus, these objects are, in a certain sense, similar to fractal objects, presenting the self-similarity property. It is important to notice that the natural objects cannot be considered true fractals once its self-similarity is manifested only at a limited number of scales. They are pseudo-fractal objects. \cite{Carlin00} demonstrates the efficiency of the modelling of objects of the real world by tools of the fractal geometry.

In problems of computational vision, an important task is the identification and classification of objects in a natural scene. In the conventional process of a computational vision system, an important step is the calculus of numbers capable of describing the analyzed object, the descriptors. When the object is modelled as a pseudo-fractal entity, an important descriptor that can be immediately obtained is the fractal dimension. The concept of fractal dimension was defined for true fractal objects and is formally defined in \cite{M68} as:

\begin{equation}
	D = \lim_{u\rightarrow0}\frac{\ln{N}}{\ln{L/u}},
\end{equation}

where $L$ is the length of the fractal object, $u$ is the length of a measure line smaller than $L$ and $N$ is the number of objects of length $u$ necessary to ``cover'' the fractal object. In the case of pseudo-fractal objects the literature shows the use of some methods based on the original concept, as the Box-counting \cite{costa-2000}, Minkowski sausage \cite{books/tricot}, mass-radius \cite{costa-2000}, and others.

In practice, fractal dimension demonstrates to be a measure of the spatial occupation of the object or still of its complexity. Any method used to estimate the fractal dimension must capture the complexity of the shape at different levels of observation (micro and macro scales). Such process elucidates the importance of the fractal dimension in the description of the geometric structure of the shape, consequently, distinguishing it from other objects, as required by the pattern recognition algorithms.

\subsection{Curvature Scale Space}

Curvature is a powerful tool in the shape analysis field. Its use enables us to study the behavior of a curve by its changes in orientation. Thus, it is possible to detect relevant points in the shape which can be used for its description and identification. The curvature is defined in terms of derivatives as

\begin{equation} 
k(t) = \frac{x^{(1)}(t)y^{(2)}(t) - x^{(2)}(t)y^{(1)}(t)}{(x^{(1)}(t)^2 + y^{(1)}(t)^2)^{3/2}},
\end{equation}
where $^{(1)}$ and $^{(2)}$ denote the first and second derivatives, respectively, and $\vec{C(t)} = (x(t), y(t))$ is a parametric vector representing the curve under analysis.

The study of the curvature property in the scale space has shown great robustness in the identification of shapes under many types of transformations, such as scaling, orientation changes, translation and even noise. The Curvature Scale Space (CSS) is achieved by applying a multi-scale transformation over the original curve. This transformation is, basically, the convolution of a Gaussian function $g(t,\sigma)$ over the parametric curve. This is performed in order to reduce the effects of noise and high frequency information before the curvature measurement \cite{costa-2000,W84,journals/tip/MokhtarianA04}:

\begin{equation}
X(t,\sigma) = x(t) \ast g(t,\sigma),
\end{equation}
\begin{equation}
Y(t,\sigma) = y(t) \ast g(t,\sigma),
\end{equation}
where
\begin{equation}
g(t,\sigma) = \frac{1}{\sigma \sqrt{2\pi}}\exp^{\frac{-t^{2}}{2\sigma^2}},
\end{equation}
is the Gaussian function with standard deviation $\sigma$ and $\ast$ denotes convolution. 

\begin{figure}[htbp]
	\centering
	\begin{tabular}{cc}
		\includegraphics[width=0.35\textwidth]{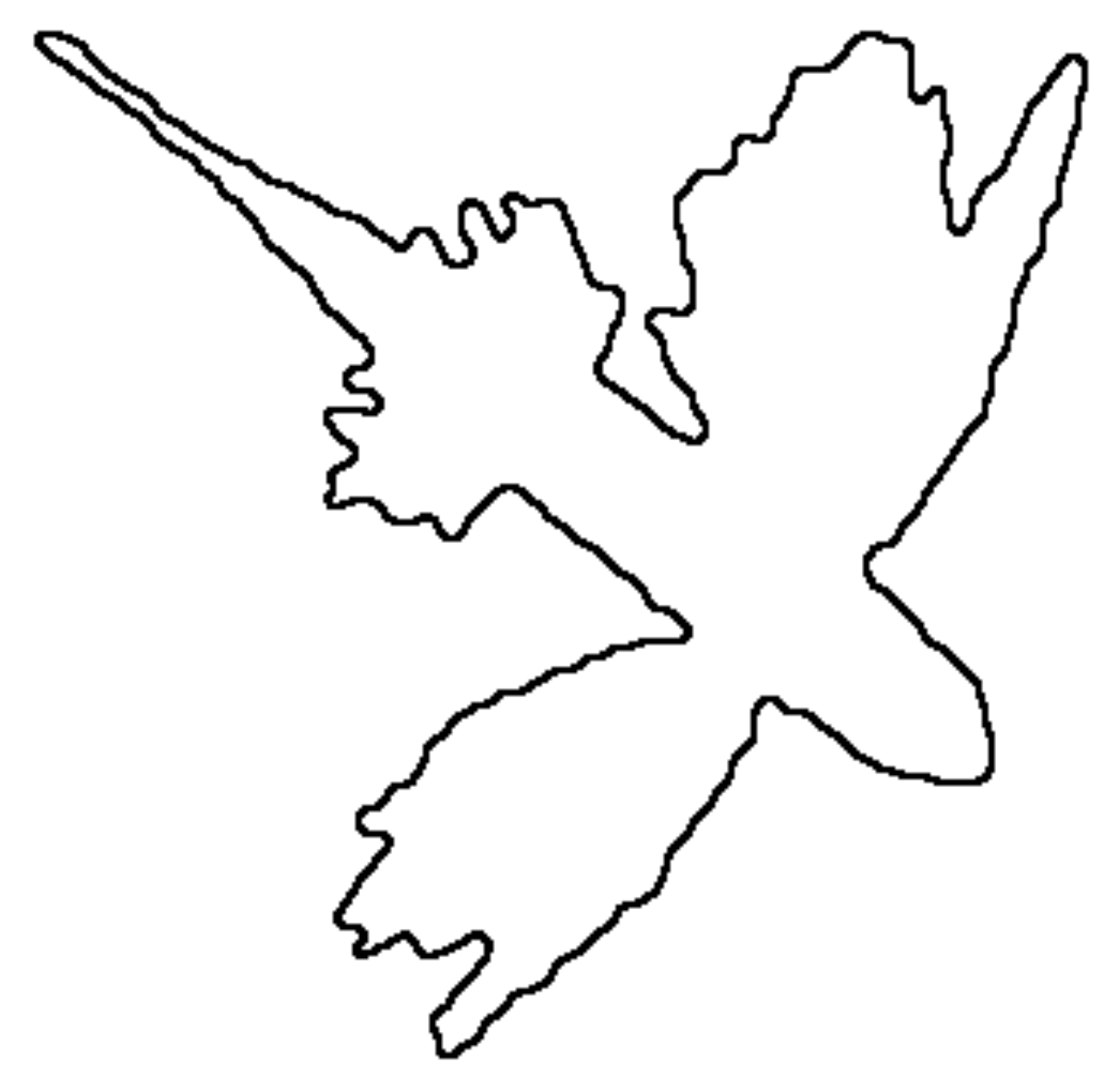} & \includegraphics[width=0.55\textwidth]{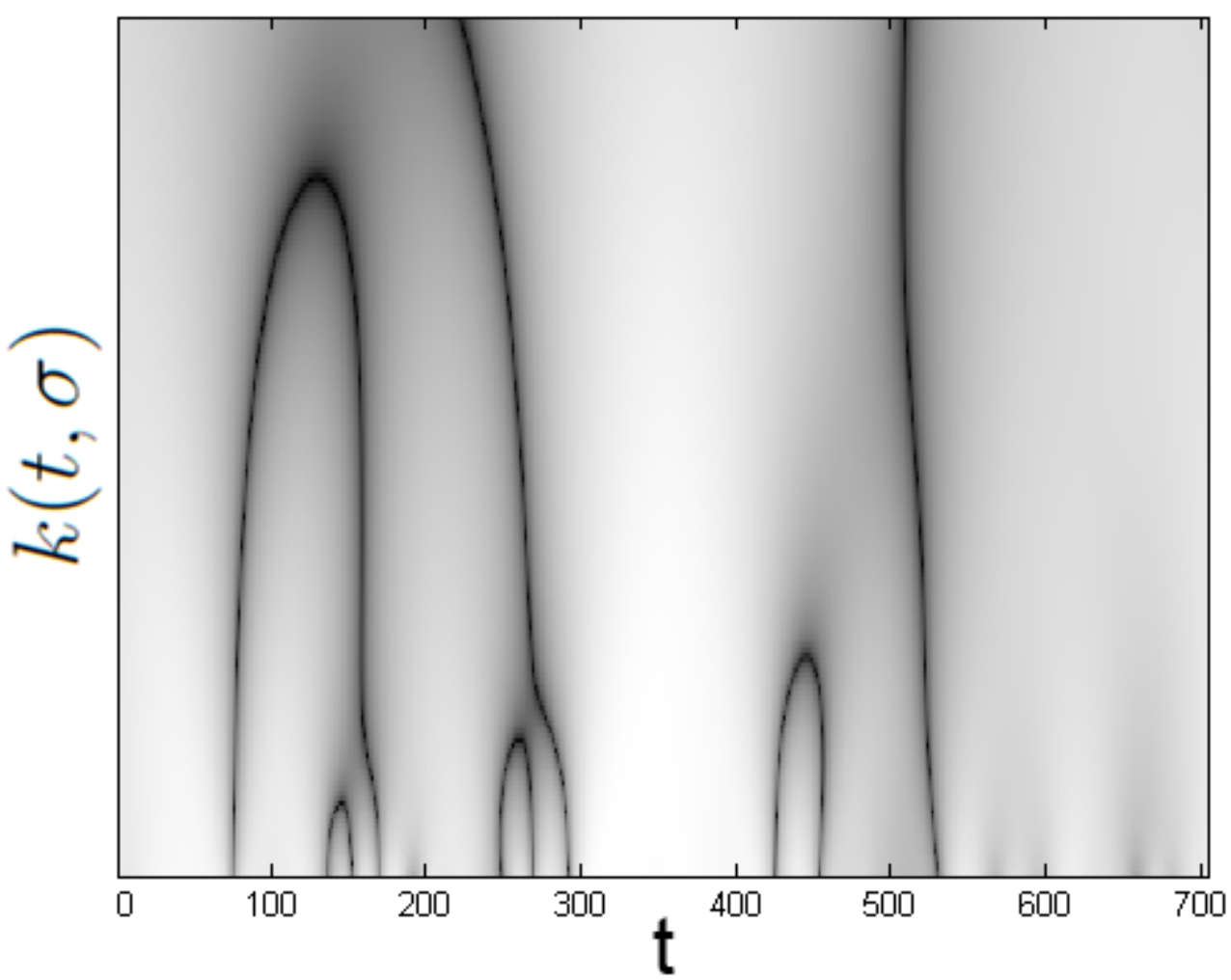}\\
		(a) & (b) \\
	\end{tabular}
	\caption{(a) Fish contours;(c) Curvature Scale Space.}
	\label{fig:css}
\end{figure}

Thus, curvature equation considering the scale evolution is defined as

\begin{equation} 
k(t,\sigma) = \frac{X^{(1)}(t,\sigma)Y^{(2)}(t,\sigma) - X^{(2)}(t,\sigma)Y^{(1)}(t,\sigma)}{(X^{(1)}(t,\sigma)^2 + Y^{(1)}(t,\sigma)^2)^{3/2}},
\end{equation}

where the value of $\sigma$ is linearly increased to achieve the set of curves, which composes the Curvature Scale Space of a given shape (Figure \ref{fig:css}). 

\subsection{Estimating fractal dimension by Curvature}

\begin{figure}[htbp]
	\centering
	\begin{tabular}{cccc}
		\includegraphics[width=0.22\textwidth]{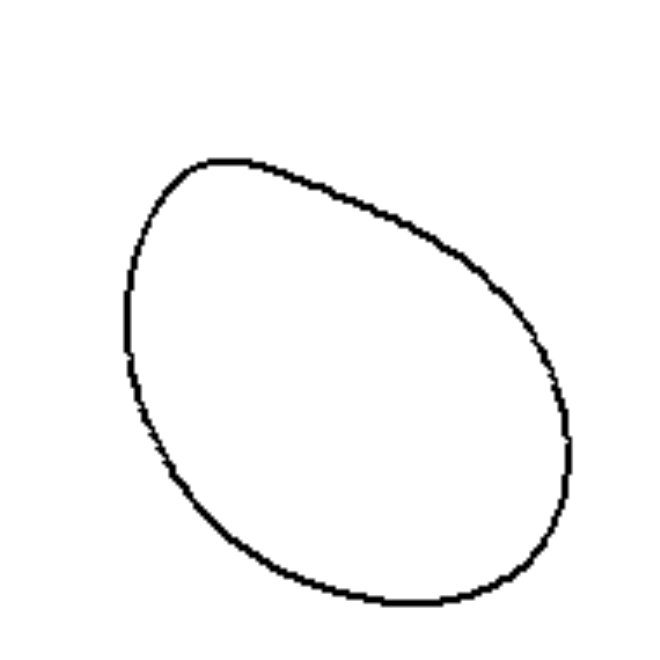} & \includegraphics[width=0.22\textwidth]{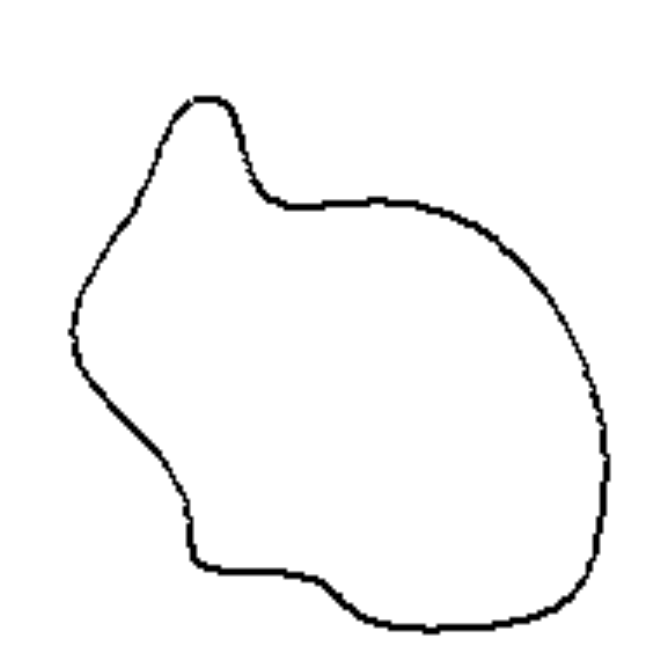} & \includegraphics[width=0.22\textwidth]{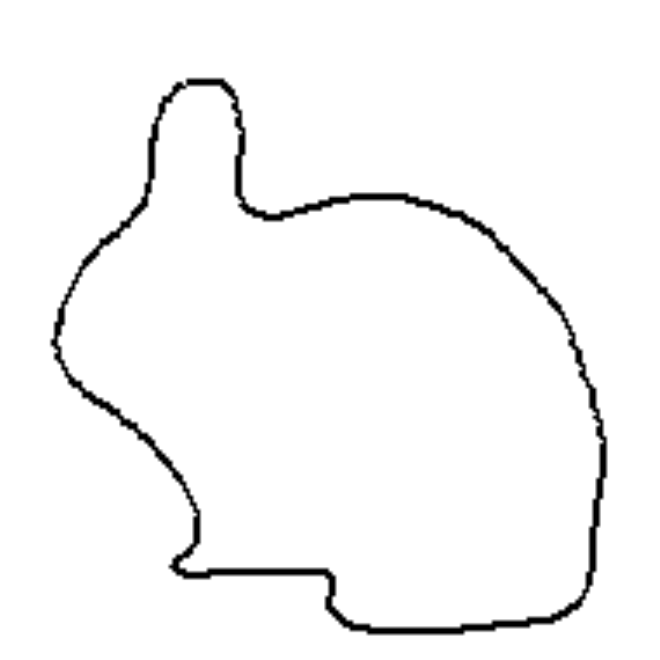} & \includegraphics[width=0.22\textwidth]{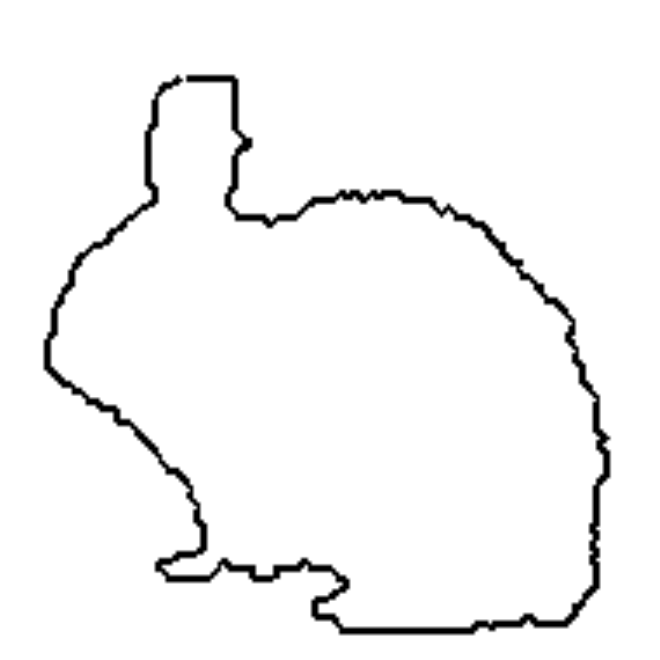} \\
		(a) & (b) & (c) & (d)\\
	\end{tabular}
	\caption{Smoothing of a shape contour as the $\sigma$ parameter changes.}%5.7183, 13.0161, 23.9403, 401.2383
	\label{fig:var_sigma}
\end{figure}

In the CSS approach, a Gaussian function is applied over the parametric curve in order to reduce the amount of high frequency information and noise present in the original shape. As the value of the $\sigma$ parameter in the Gaussian function decreases, the higher is the smoothing level applied over the curve. During this process, the aspect of the shape changes. We note that the higher changes in the contour orientation are diminished due to the reduction of the information contained in the curve. As a result, the shape becomes more similar to a circle (Figure \ref{fig:var_sigma}). This process reflects in the complexity of the shape, so that, curvature scale space can be used to estimate the fractal dimension of a curve.

\begin{figure}[htbp]
	\centering
	\begin{tabular}{cc}
		\includegraphics[width=0.42\textwidth]{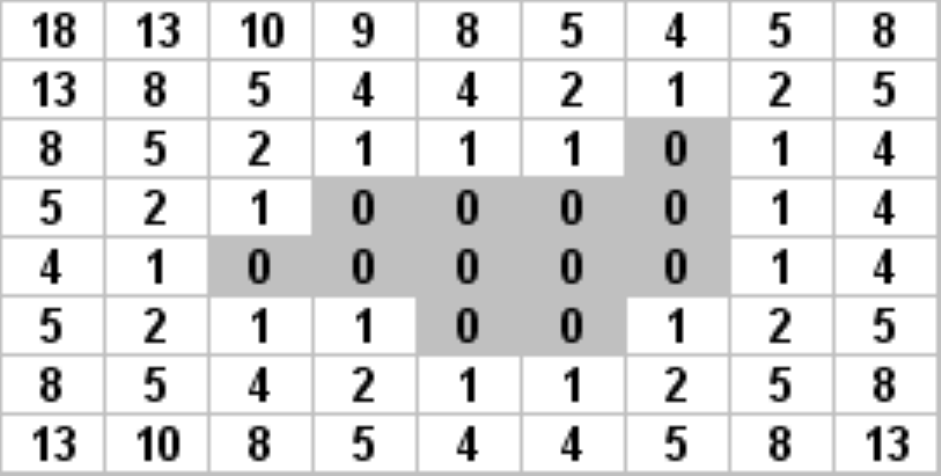} & \includegraphics[width=0.42\textwidth]{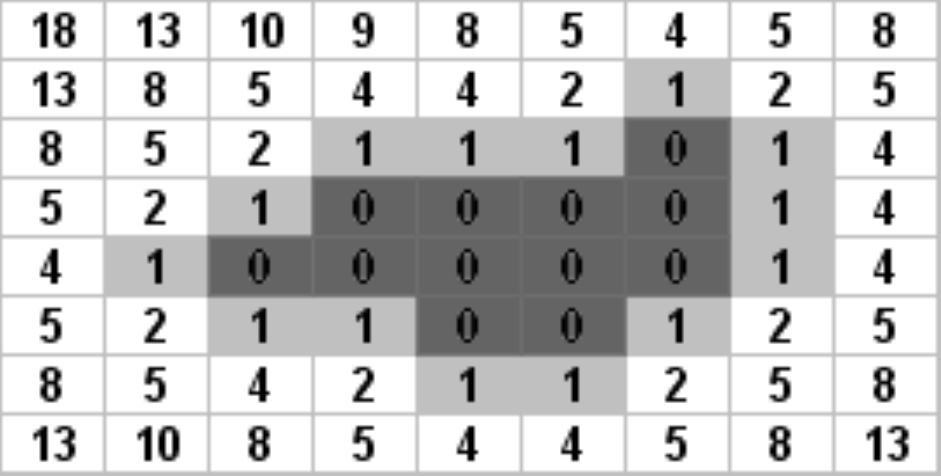} \\
		(a) & (b) \\
		\includegraphics[width=0.42\textwidth]{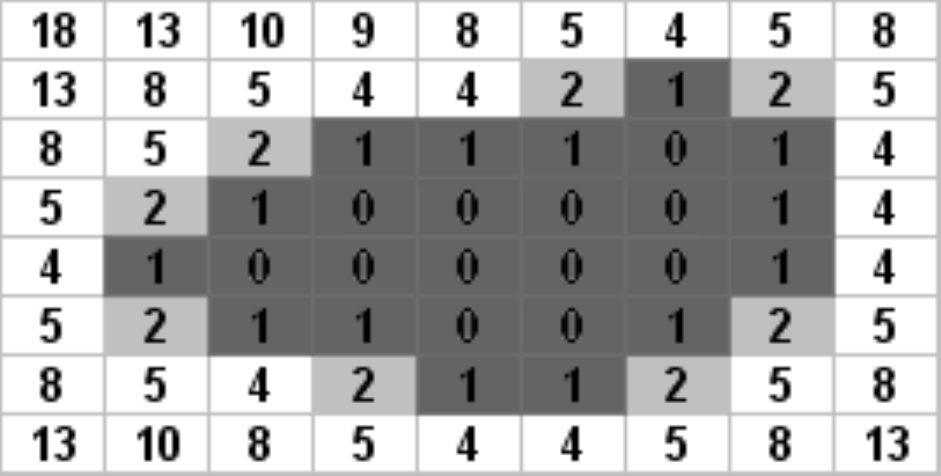} & \includegraphics[width=0.42\textwidth]{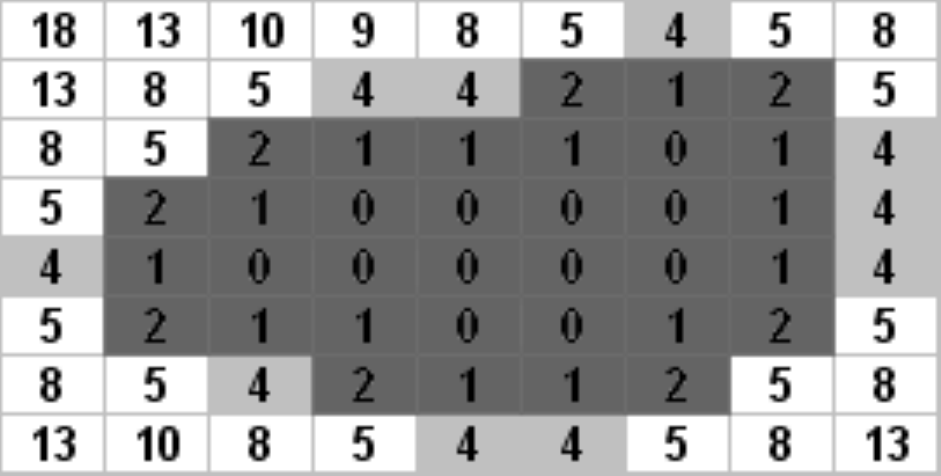} \\
		(c) & (d)\\
	\end{tabular}
	\caption{Process of shape dilation by the Bouligand-Minkowski method (light gray, points correspondent to the current radius $r$; dark gray, accumulated area): (a) Original shape ($r = 0$); (b) $r^2 = 1$; (b) $r^2 = 2$; (b) $r^2 = 4$.}
	\label{fig:var_dilat}
\end{figure}

In the Bouligand-Minkowski method, the shape of an object is dilated and its area computed for different dilatation radius $r$ to estimate the shape complexity (Figure \ref{fig:var_dilat}). A similar idea can be applied to the curvature. Like in the Bouligand-Minkowski method, we define the amount of curvature $S(\sigma)$ as the sum of the module of all curvature coefficients computed for a shape:

\begin{equation}
S(\sigma) = \sum_{t} \left| k(t,\sigma) \right|,
\end{equation}

where $S(\sigma)$ represents the resulting orientation of the curve at scale $\sigma$. For each $\sigma$ value used, a different amount of curvature $S(\sigma)$ is computed. However, defining a set of $\sigma$ values to be used is a difficult task. Thus, as in CSS method, we opt to linearly increase its value. 

By accumulating the subsequent values for $S(\sigma)$, it is possible to create a relation of dependence among scales, where posterior scales are influenced by previous scales. This is just similar to the process of dilating a shape performed by the Bouligand-Minkowski method, where a dilatation using a radius $r$ is the accumulation of every point at a distance from the shape ranging from $0$ to $r$. Thus, the accumulated amount of curvature is defined as

\begin{equation}
A(\sigma) = \sum_{i=0}^{\sigma} S(i).
\end{equation}

\begin{figure}[htbp]
	\centering
	\begin{tabular}{cc}
		\includegraphics[scale=0.35]{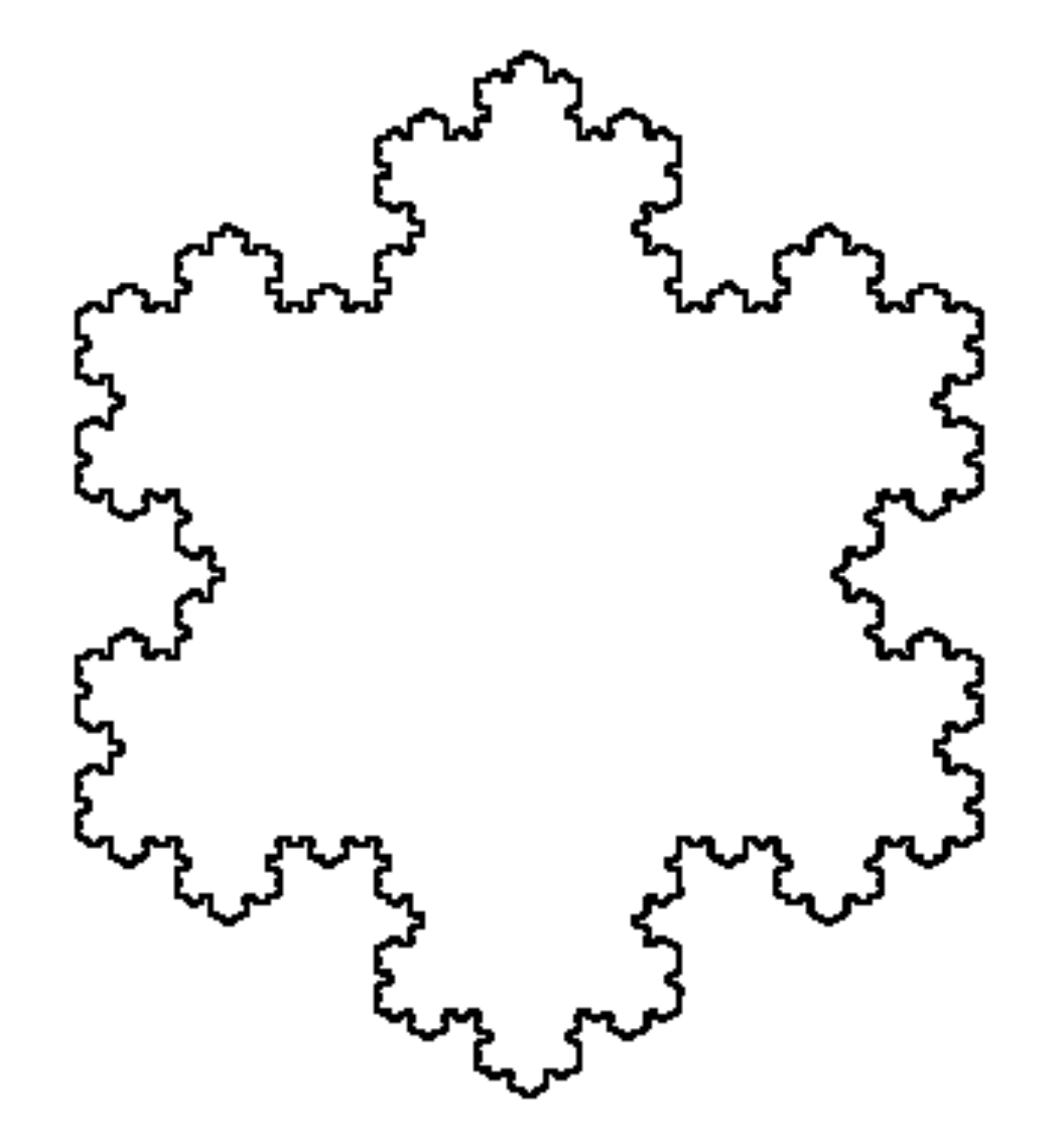} & \includegraphics[scale=0.35]{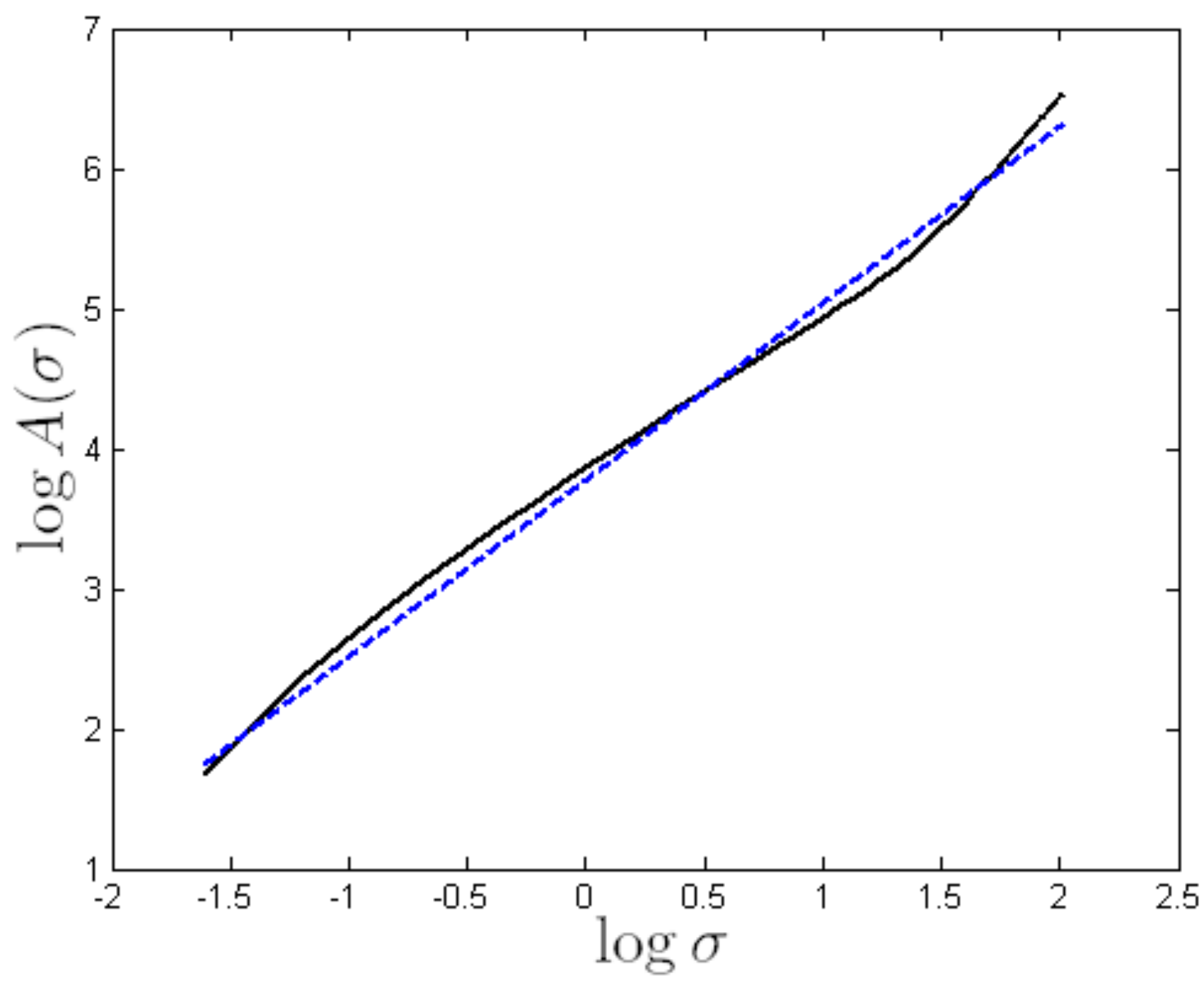} \\
		(a) & (b)\\		
	\end{tabular}
	\caption{(a) Koch snowflake ($d = 1.2618$); (b) Curvature Fractal Dimension ($d = 1.2644$).}
	\label{fig:loglog}
\end{figure}

The degree of space filling of a curve can be quantified by considering its order of growth. For this, it is necessary to analyze the behavior of $\sigma \times A(\sigma)$ curve in the logarithmic scale. Figure \ref{fig:loglog} shows the log-log plot of the $\sigma \times A(\sigma)$ curve computed for the Koch snowflake (or Koch star). The Koch snowflake is a mathematical curve of infinite length and one of the earliest fractal curves described in the literature. At each iteration, one line segment of this curve is replaced by four line segments, each one having one-third of the size of its previous stage, what leads to the fractal dimension $d = \log{4}/\log{3} \cong 1.2618$. Note that the curve achieved obeys a power law. According to \cite{books/tricot}, this power law can be used to estimate the fractal dimension of the original shape:

\begin{equation} 
d = \lim_{\sigma \to 0} \frac{\log{A(\sigma)}}{\log{\sigma}}.
\end{equation}

By applying a line regression over the log-log plot of $\sigma \times A(\sigma)$, it is possible to estimate a straight line with slope $\alpha$, where $d = \alpha$ is the estimated fractal dimension for the closed curve $\vec{C(t)}$. Notice that the method estimated the fractal dimension with $d = 1.2644$. This result is very close to the theoretical fractal dimension of the Koch snowflake and shows the accuracy of the proposed method and its straight relation to the fractal theory.  

\section{Multi-scale Fractal Dimension}

As previously described, the changes in the value of the $\sigma$ parameter in the Gaussian function affects the aspect of the shape (Figure \ref{fig:var_sigma}). As a result, small variations are incorporated to the log-log curve $\sigma \times A(\sigma)$. Thus, eventually, a single non-integer number $d$ may be not adequate to characterize all the complexity richness existent in an object. Moreover, non-fractal objects, such as a shape curve, has finite size. This implies that its dimension goes to zero as the visualization scales increase. This is corroborated by the tendency of the shape to become more similar to a circle as the the information of high frequencies is reduced.

\begin{figure}[htbp]
	\centering
	\begin{tabular}{cc}
		\includegraphics[scale=0.35]{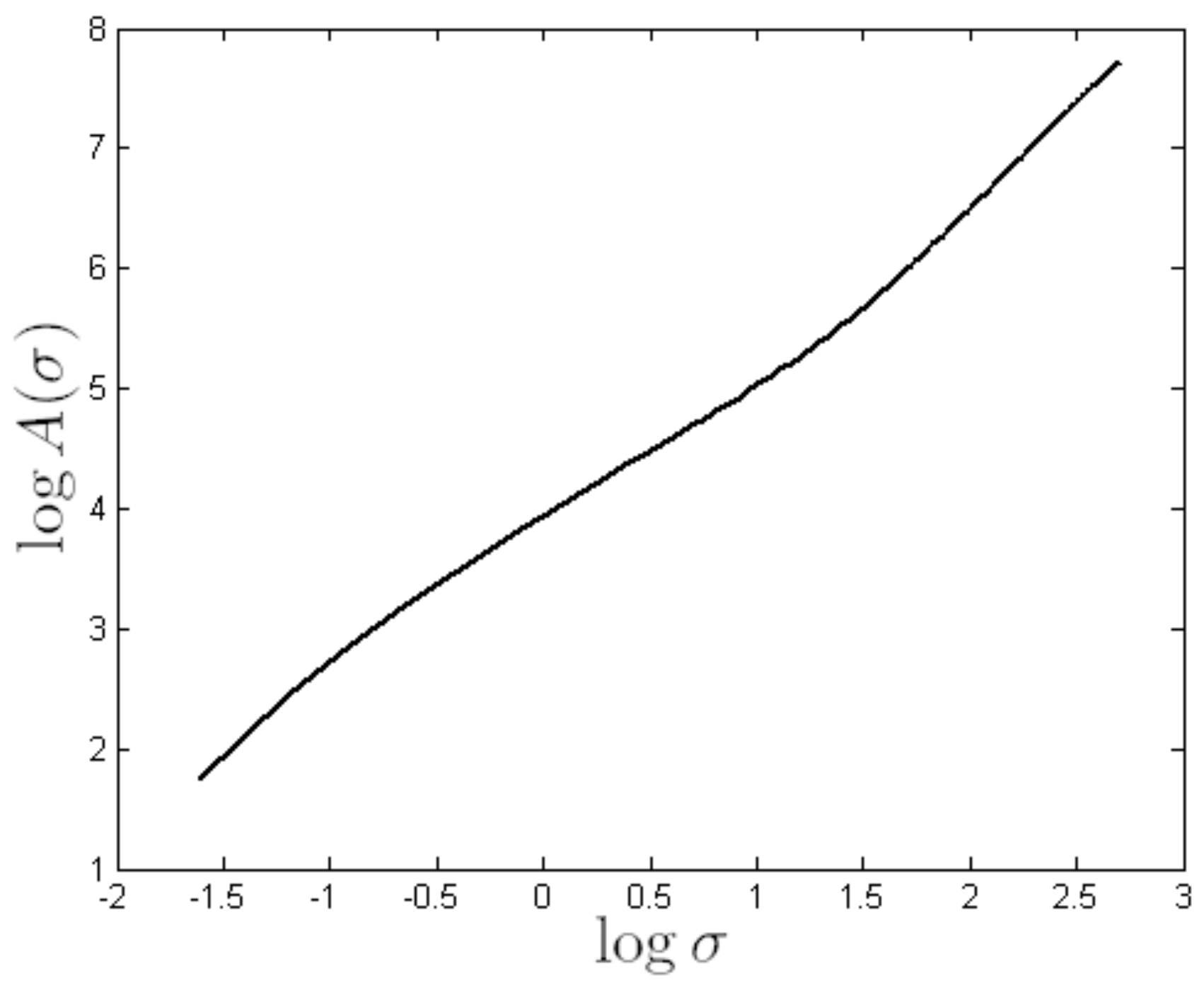} & \includegraphics[scale=0.35]{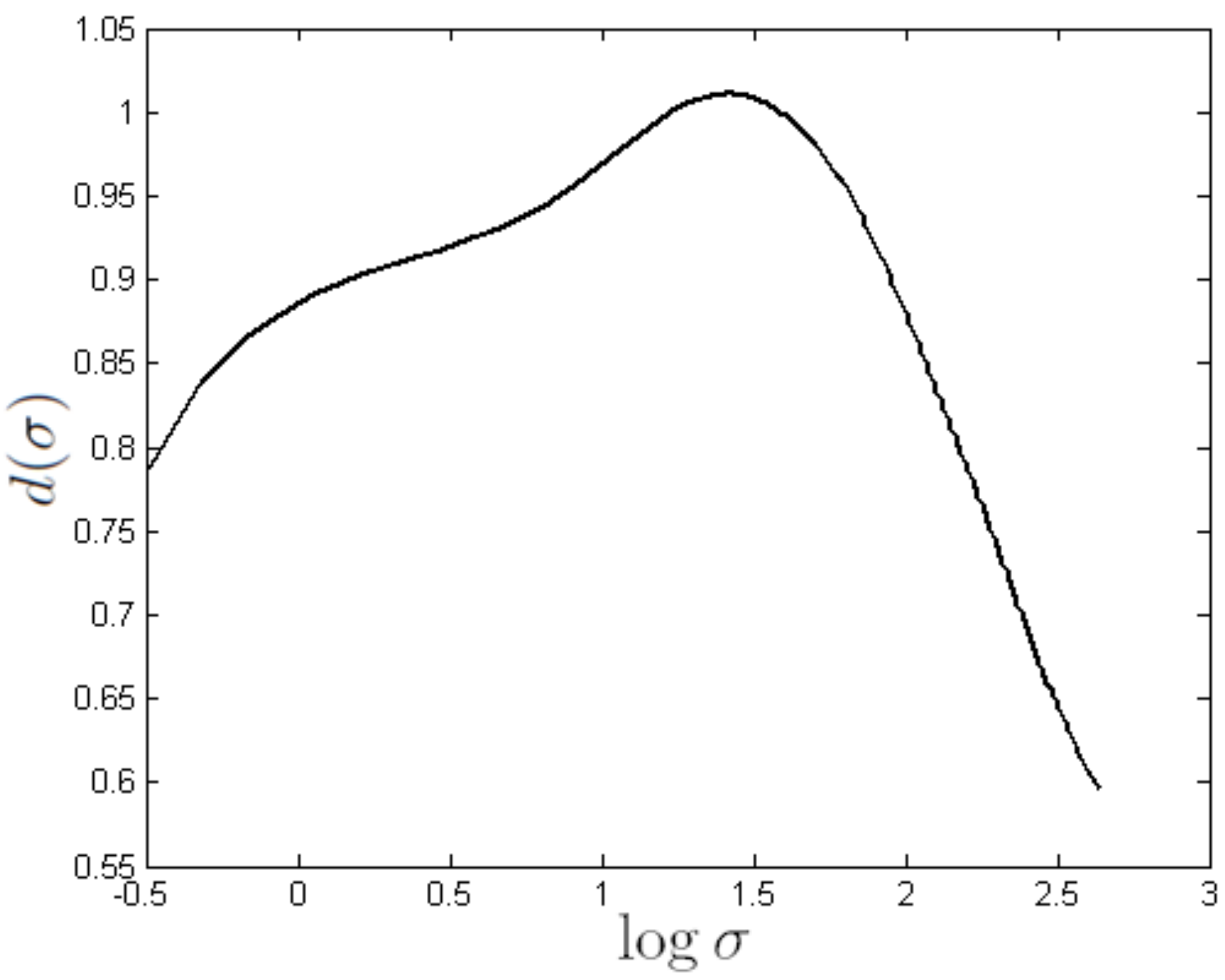}\\		
		(a) & (b)\\		
	\end{tabular}
	\caption{(a) log-log curve. (b) Multi-scale Fractal Dimension.}
	\label{fig:multiescala}
\end{figure}

In order to solve this deficiency of the fractal dimension methods and aiming to provide a better description of objects in terms of complexity, the Multi-Scale Fractal Dimension has been proposed \cite{bb21635,GW2002,Plotze-2005}. This approach exploits the infinitesimal limit in the linear interpolation by using the derivative, thus yielding a function that binds the changes in the object complexity to the visualization scale changes (Figure \ref{fig:multiescala}). This function enables us to perform a more effective discrimination of the object and it is defined as:

$$d(\sigma) = \frac{d\log{A(\sigma)}}{d \log{\sigma}},$$
where $d(\sigma)$ represents the complexity of the object at scale $\sigma$.

\section{Experiments}

\begin{figure}[ht]
	\centering
		\includegraphics[width=0.8\textwidth]{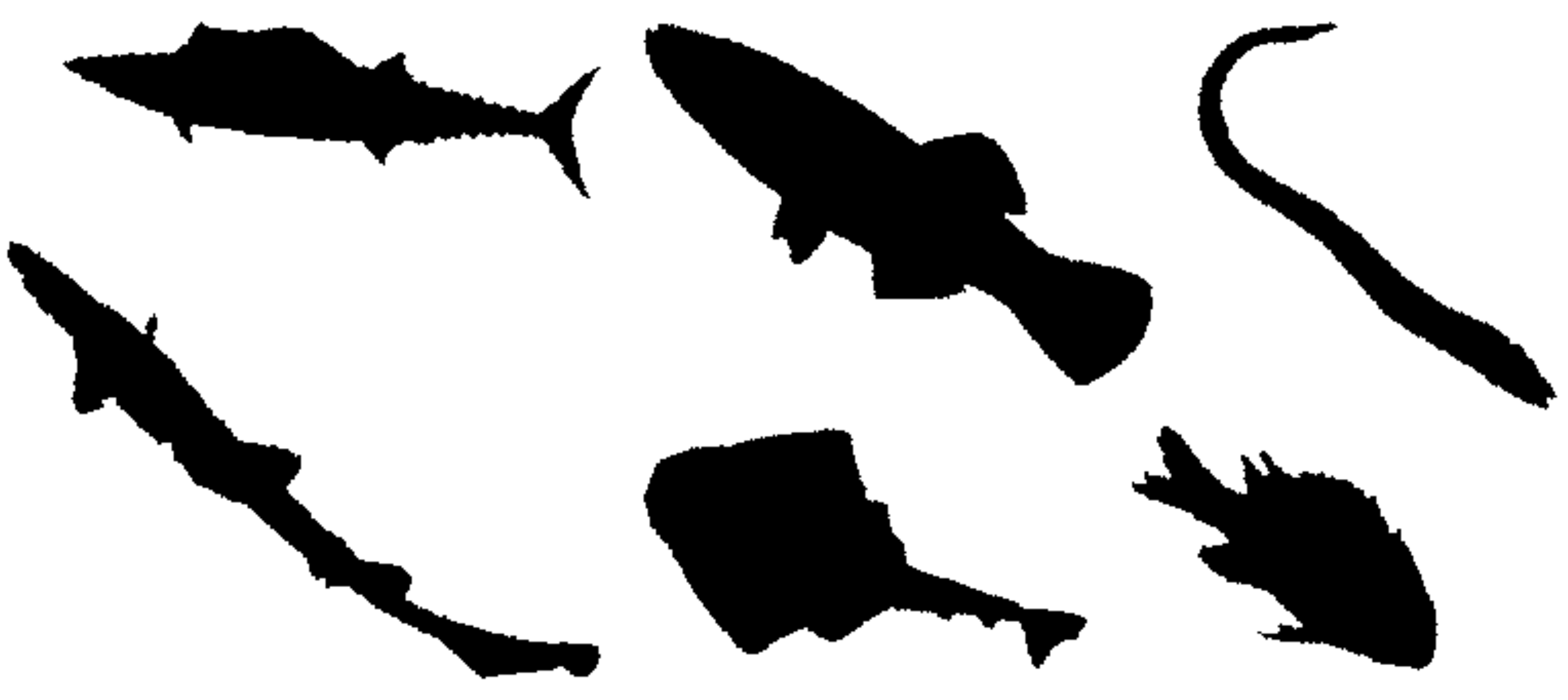}
	\caption{Examples of fish images used in the experiments.}
	\label{fig:FishSamples}
\end{figure}

An experiment was performed to evaluate the proposed fractal approach as a feasible shape descriptor. The experiment aimed the classification of a large database composed by shapes under rotation and scale transformations. The database consists of 1100 classes, where each class is one of the 1100 fish contours obtained from the database available at \cite{fishdatabaseURL}. Each class is composed of 10 different manifestations (5 rotations and 5 scales) of the fish contour associated, thus resulting in a database containing 11000 shapes. Figure \ref{fig:FishSamples} shows some examples of fish contours present in the database.

Statistical evaluation of the Multi-scale curves were carried out using the Linear Discriminant Analysis (LDA). The LDA is a supervised statistical classification method \cite{everitt-2001,fukunaga-1990}, which searches a linear combination of the descriptors (independent variables) that results in its class (dependent variable). Thus, the LDA method attributes an observation $x$ to the class $i$ which presents the higher conditional probability:

\begin{equation}
	P(i|x) > P(j|x), \forall j \neq i.
\end{equation}

where $P(i|x)$ is the conditional probability of $i$ given $x$. Considering that $P(i|x)$ obeys a multivariate gaussian distribution with a single covariance matrix for all the classes, an object is attributed to the class $i$ which provides the higher value for the function $f_{i}$:

\begin{equation}
	f_{i} = \mu_{i}C^{-1}x_{k}^{T}-0.5\mu_{i}C^{-1}x_{k}^{T}+\log(p_{i}),
\end{equation}

where $\mu_{i}$ is the mean of the descriptors of class $i$, $C$ is the covariance matrix of the data set and $p_{i}$ is the \emph{a priori} probability of class $i$.

The statistical analysis was carried out over the samples using a K-fold cross-validation scheme with 5 folds. In order to provide a fair evaluation of the properties of the proposed method (such as the tolerance to rotation and scaling), one manifestation of rotation and scaling was considered in each class, in each fold.

\section{Results and Discussion}

To provide a robust evaluation of the proposed approach, a comparison with the Bouligand-Minkowski method was considered \cite{books/tricot,Plotze-2005,bruno-2008}. The Bouligand-Minkowski is one of the most accurate methods to estimate the fractal dimension. In this method, the shape of an object is dilated by a disc of radius $r$ and the influence area is used to estimate the shape complexity. The influence area is very sensitive to structural changes of the shape, thus even small changes in the shape structure can be detected. For this experiment, we considered a dilation radius $r = 50$.

Finite Difference method \cite{IB-D85044} was applied over log-log curves computed for both Bouligand-Minkowski and the proposed approach, thus resulting in a Multi-scale fractal dimension curve characteristic for that shape. A total of 50 descriptors was considered from Multi-scale curves for shape characterization.

Table \ref{tab:results} presents the classification results yielded for each approach. Variations in the size or orientation of a shape are a very common issue in shape analysis. Tolerance to these transformations is a fundamental characteristic for a method to be considered robust in shape analysis applications. As we can see, the Bouligand-Minkowski method is not scale invariant. Thus, in order to reduce this problem and to perform a fairer comparison, a second experiment was performed where the shapes were first normalized according to their diameter \cite{Torres2003}.

\begin{table}[!ht]
	\caption{Comparison results for different complexity estimation methods.}
	\centering		
		\begin{tabular}{ c c c }
		\hline
		Method & Samples correctly classified & Success rate\\				
		\hline
		Proposed approach & 10685 & 97.14 \\
		Bouligand-Minkowski & 1573 & 14.30 \\		
		Bouligand-Minkowski (normalized)& 8807 & 80.06 \\		
		\hline
		\end{tabular}
		\label{tab:results}
\end{table}

Both methods compared are tolerant to orientation changes in the shape. Nevertheless, it is important to emphasize that images exist only in the discrete space, and they cannot be freely rotated. A small error is always added to the multi-scale curve due to shape rotation. The tolerance in the Bouligand-Minkowski method is due to the use the Euclidean distance during the computing of the influence area. This is a distance measure that is not affected by transformations in orientation. In our proposed approach, the rotation tolerance is most since curvature is not sensitive to orientation changes in the shape. The curvature is a measure related to the curve orientation at a specific point, i.e., the changes in the direction inside the curve. Thus, the changes in the orientation of the shape do not affect the orientation among its inner segments.

\begin{figure}[htbp]
	\centering
	\includegraphics[width=0.8\textwidth]{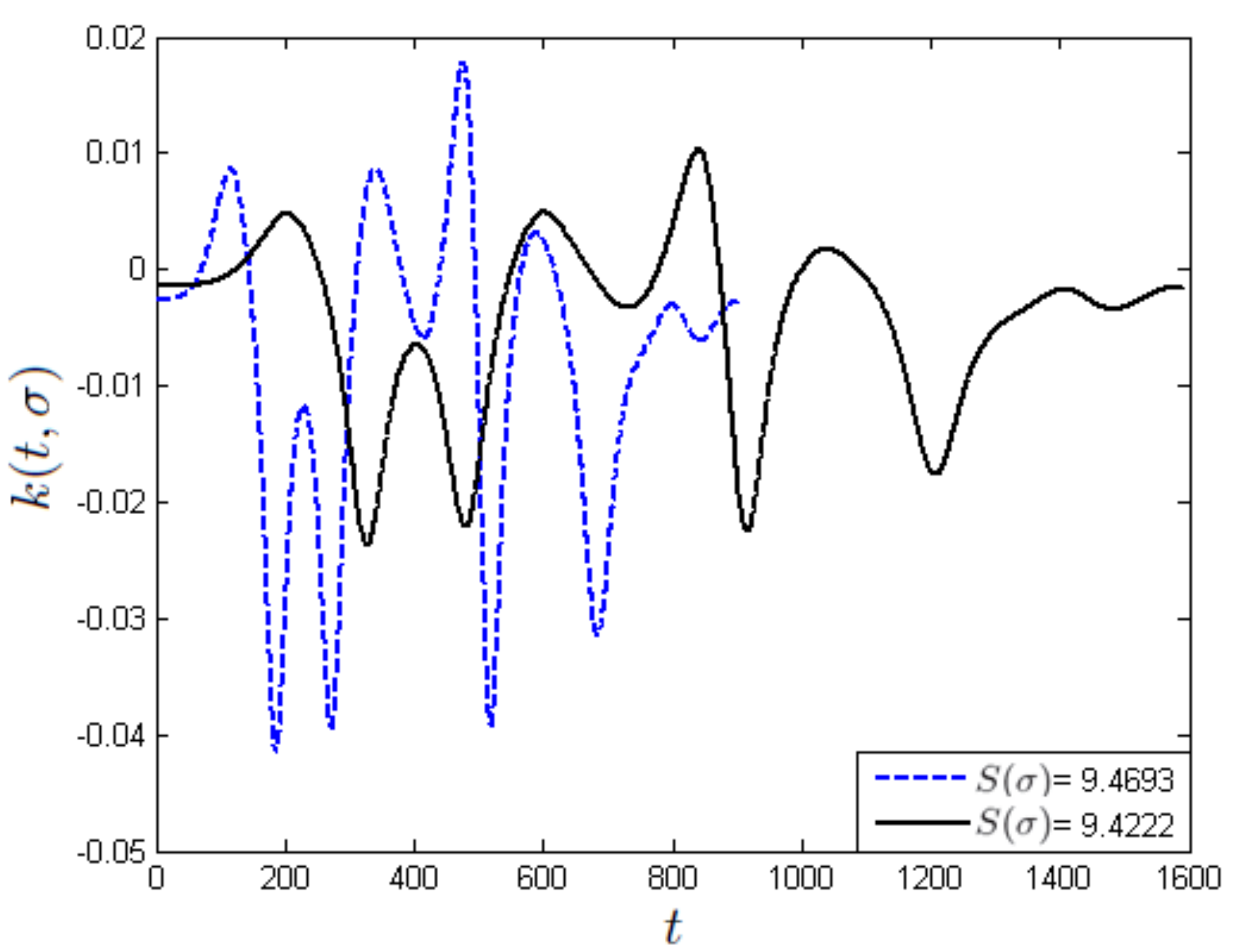}
	\caption{Curvature and the sum of its coefficients, $S(\sigma)$, for two contour at different scales.}
	\label{fig:escala}
\end{figure}

Unlike the Bouligand-Minkowski method, which needs a normalization step performed directly over the shape image, the proposed approach is tolerant to scale changes and so no normalization of the data is necessary. The main reason for its scaling tolerance is the use of the sum of the module of the curvature coefficients as the basis to estimate the shape complexity (Figure \ref{fig:escala}). As in the rotation tolerance, scaling changes does not affect the orientations of the inner structures of the contour. It only increases when the size of the shape increases, i.e., the number of points in its contour. Minimum and maximum local points are also preserved when the direction changes in the contour as well as the sum of its coefficients. Thus, the behavior of the multi-scale fractal dimension curve is preserved even for scaled shapes. % The same is not valid for the Bouligand-Minkowski method, where the influence area depends on a dilation radius which depends on the shape size.

\section{Conclusion}
\label{sec:conclusion}
This work presented the study of a novel technique for extracting shape descriptors using a fractal methods based on the curvature scale space concept.

Results demonstrated that the proposed technique achieved a great accuracy in the classification of the shape database containing rotated and scaled samples, showing an interesting correctness rate of 97.14\%. The method also showed the best accuracy in comparison to the Bouligand-Minkowski method, as it is being intrinsically invariant to scale, a property which cannot be found in the conventional Bouligand-Minkowski method. 

As the Bouligand-Minkowski method is shown in the literature as an efficient technique to obtain fractal shape descriptors, the results presented in this work suggest that the proposed technique is a worthy option for providing shape descriptors for classification tasks, as it is invariant to rotation and scale, operations commonly found in practical situations. Such fact turns possible the application of the proposed technique to various problems in computer vision involving the classification of shapes.

We still suggest, as a future work, a deeper study about this novel method, in order to verify its application to another pattern recognition and computer vision problems.

\section*{Acknowledgments}

A.R.B. acknowledges support from FAPESP (2006/54367-9).
J.B.F. acknowledges support from CNPq (870336/1997-5) and FAPESP (2006 / 53959-0).
O.M.B. acknowledges support from CNPq (Grant \#308449/2010-0 and \#473893/2010-0) and FAPESP (Grant \# 2011/01523-1). 

%\bibliography{Bibliografia}

%merlin.mbs 2010-03-15 4.21a (PWD, AO, DPC)
%Control: key (0)
%Control: author (8) initials jnrlst
%Control: editor formatted (1) identically to author
%Control: production of article title (0) allowed
%Control: page (1) range
%Control: year (1) truncated
%Control: production of eprint (0) enabled
%

\end{document}